\documentclass[magazine,twocolumn]{IEEEtran}
\usepackage{amsmath,amsfonts}
\usepackage{algorithmic}
\usepackage{algorithm}
\usepackage{array}
\usepackage[caption=false,font=normalsize,labelfont=sf,textfont=sf]{subfig}
\usepackage{textcomp}
\usepackage{stfloats}
\usepackage{url}
\usepackage{hyperref}
\usepackage{verbatim}
\usepackage{graphicx}
\usepackage{xcolor}
\usepackage{cite}
\usepackage{comment}
\usepackage{soul}

\begin{document}

\title{Enabling Next-Generation Cloud-Connected Bionic Limbs Through 5G Connectivity}

\author{
\IEEEauthorblockN{Ozan Karaali\IEEEauthorrefmark{1}, Hossam Farag\IEEEauthorrefmark{1}, Strahinja Do\v{s}en\IEEEauthorrefmark{2} and \v{C}edomir Stefanovi\'{c}\IEEEauthorrefmark{1}}\\
\IEEEauthorrefmark{1}Department of Electronic Systems, Aalborg University, Denmark\\
\IEEEauthorrefmark{2}Department of Health
Science and Technology, Aalborg University, Denmark\\
Email: \{ozank,hmf,cs\}@es.aau.dk, sdosen@hst.aau.dk
}
\maketitle
\begin{abstract}
    Despite the recent advancements in human-machine interfacing, contemporary assistive bionic limbs face critical challenges, including limited computational capabilities, high latency, and unintuitive control mechanisms, leading to suboptimal user experience and abandonment rates. Addressing these challenges requires a shift toward intelligent, interconnected solutions powered by advances in Internet of Things systems, particularly wireless connectivity and edge/cloud computing. 
    This article presents a conceptual approach to transform bionic limbs by harnessing the pervasive connectivity of 5G and the significant computational power of cloud and edge servers, equipping them with capabilities not available hitherto. The system leverages a hierarchical distributed-computing architecture that integrates local, edge, and cloud computing layers.
    Time-critical tasks are handled by a local processing unit, while compute-intensive tasks are offloaded to edge and cloud servers, leveraging the high data rate, reliable and low latency capabilities of advanced cellular networks.
    We perform a proof-of-concept validation in a 5G testbed showing that such networks are capable of achieving data rates and fulfilling latency requirements for a natural prosthetic control, allowing for offloading of compute-intensive jobs to the edge/cloud servers.
    This is the first step towards the realization and real-world validation of cloud-connected bionic limb systems.
\end{abstract} 


\section{Introduction}

\IEEEPARstart{R}{obotic} bionic limbs (prostheses and exoskeletons) can be used to restore missing motor functions lost due to an injury or disease of the human sensorimotor systems. Due to the rapid development of technology, modern bionic limbs are advanced mechatronic systems with multiple functions, e.g., contemporary robotic prosthetic hands feature individually controllable fingers and multi-degrees-of-freedom active wrist. However, despite these developments, bionic limbs are still often rejected by their users, and one of the main reasons is the lack of efficient and user-friendly control \cite{salminger_current_2022}. 

For many years, the commercial standard in prosthetics was the 2-channel electromyography (EMG) control interface. Here, the electrical activity of two muscles responsible for opening and closing the hand before amputation is mapped to the opening and closing of the prosthesis. This is an intuitive approach suitable only when controlling a simple prosthesis with a single degree of freedom. Recently, machine learning (ML) based control schemes became commercially available. In this case, EMG signals are recorded from multiple muscles, pattern classification is used to recognize a gesture the user wants to perform, and the robotic hand is then activated to produce that gesture \cite{_myoplus_}. Although this is a big improvement with respect to the 2-channel approach, the cognitive burden of control is still fully on the user, who often needs to produce several explicit commands to perform a simple action (e.g., rotate wrist, open hand, grasp). 

Some promising methods that can improve performance and user experience were presented in the literature, e.g., the use of musculoskeletal modeling, deep learning, and/or shared control \cite{marinelli_active_2023}. In the latter approach, bionic limbs are equipped with additional sensors, like cameras or depth sensing units, so that the artificial controller detects and analyzes the environment (e.g., target objects) to implement some actions automatically (e.g., preshapes the hand and grasps the object). As demonstrated experimentally, this approach can improve performance while decreasing cognitive effort, especially when controlling complex devices \cite{mouchoux_artificial_2021}. However, it requires significant computational resources that normally are not available in bionic limbs, which are compact wearable systems.
To overcome this challenge, we recently proposed a new generation of bionic limbs that are empowered by connectivity so that they can exploit the virtually unlimited processing and storage capacity of the cloud/edge infrastructure \cite{chiariotti_future_2024}. A cloud-connected bionic limb can offload complex processing tasks to powerful edge/cloud servers, significantly alleviating the computational burden on the device itself, as well as enabling novel functionalities. This way, bionic limbs can be integrated into the Internet of Things (IoT) ecosystem, promoting important use cases in the domain of digital health. However, computational offloading demands a robust communication infrastructure capable of supporting high-speed, reliable, and low-latency data exchanges between the connected device and external servers. 

In this article, we introduce a conceptual framework for cloud-connected bionic limbs that distributes processing across local, edge, and cloud resources to enhance device capabilities. The framework integrates multiple sensory inputs (EMG, tactile sensing, motion sensors, and cameras) and command streams with a three-tier processing system: local units handle time-critical basic control, edge servers process complex tasks like computer vision and context analysis that still need to run online, and cloud infrastructure manages resource-intensive operations like model training and data analytics, which are not time-critical. The approach has the potential to address the key challenges in the control of bionic limbs, including local computational constraints, power efficiency, and the need for sophisticated environmental perception to run advanced control algorithms and for the timely closing of the control loop. Through experimental evaluation using a 4G/5G testbed, we demonstrate the feasibility of offloading compute-intensive tasks while maintaining responsive control within the required latency bounds, which is a crucial preliminary step in determining the feasibility of cloud-connected bionic limb systems. 

This article is organized as follows. Section ~\ref{sec:architecture} presents the system architecture. Section ~\ref{sec:evaluation} is devoted to the experimental evaluation of the cellular connectivity for the closed-loop control of the bionic limb in the edge.
Section ~\ref{sec:conclusion} concludes the paper.

\section{System Architecture}
\label{sec:architecture}

\begin{figure}[!t]
    \centering
    \includegraphics[width=\columnwidth]{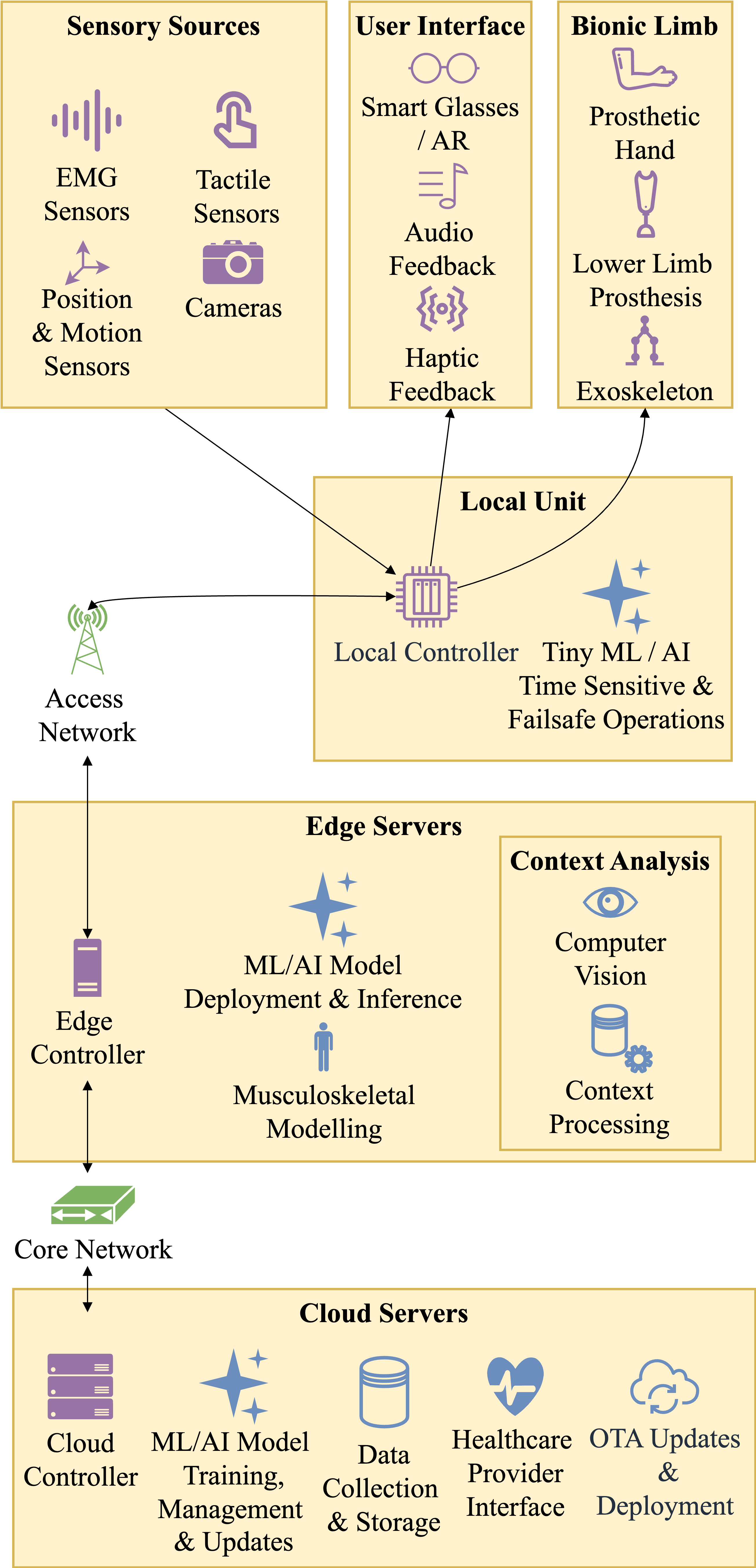}
    \caption{System architecture for the control of connected bionic limbs showing data flow between sensors, edge computing, and cloud servers.}
    \label{fig:system_arch}
\end{figure}

\autoref{fig:system_arch} illustrates the proposed cloud-connected assistive system architecture.
It integrates sensory data sources, command streams and feedback interfaces with a distributed processing framework that comprises local, edge, and cloud servers performing control and monitoring functions. The sensed information is transmitted to the edge and cloud infrastructure, where dedicated servers (controllers) compute commands and feedback signals that are transmitted back to the device and the user, respectively.

The system, therefore, operates in a shared control framework, with controllers at various levels of hierarchy detecting the user's motion intents and aiding the user to perform the motions leveraging the sensed information (e.g., camera feeds).
It is of paramount importance that the user gets an intuitive sense of agency when using the bionic limb:
In this respect, it was demonstrated that a comfortable use of a bionic limb requires response times between 100-125 milliseconds~\cite{farrell_optimal_2007} to avoid user dissatisfaction and provide accurate grip control.
In an ideal scenario, a local controller attached to the bionic limb would perform the required computations; however, this is virtually impossible in practice.
Even if the control could be executed locally within the latency budget, which could hardly be done due to hardware constraints, the energy required for running the control logic would quickly deplete the device's battery.
For instance, simple grasping categorization algorithms require significant resources, e.g., NVIDIA Jetson TX2 uses 5-11W of power for this operation alone, which will drain the standard battery in a matter of hours~\cite{ragusa_video_2021}.
On the other hand, the indicated latency budget opens the possibility of offloading the computations to powerful edge and cloud servers, which, in turn, enables the implementation of advanced control and monitoring functions.
Specifically, complex tasks such as computer vision and deep learning can run on edge/cloud servers, promoting local energy efficiency while allowing access to large, remotely located processing resources.
Related works support this concept, showing that offloading computationally intensive tasks to the edge can both reduce latency of closing the control loop and dramatically reduce energy consumption at the edge device, see for instance~\cite{10.1145/3615991.3616403}.

The local, edge, and cloud layers are responsible for handling processing tasks on progressively longer time scales and of increasing complexity.
This framework natively integrates an AI/ML workflow in which:
\begin{itemize}
    \item Cloud servers handle data collection, model training, management, and updates, which happen in non-real time. 
    \item Edge servers manage model deployment and inference in real-time.
    \item Local units execute time-critical and fail-safe operations.
\end{itemize}

The critical communication links reside in the wireless access network connecting local controllers and edge servers, as they are tasked with providing high-throughput, reliable, and low-latency services facilitating the closing of the real-time control loop; this is further investigated in Section~\ref{sec:evaluation}.
On the other hand, edge-to-cloud communication occurs through the core network; this connection can be in non-real time, facilitating offline data collection for model training, management, and updates.  

In the rest of the section, we briefly describe the main components of the system.

\subsection{Smart bionic limbs}

\subsubsection{Mechatronics}

The proposed conceptual framework may be used with a variety of smart bionic limbs \cite{chiariotti_future_2024}. Robotic prosthetic hands feature different mechatronic solutions driven by a trade-off between robustness and power versus dexterity. For instance, in the Michelangelo Hand (Ottobock) all fingers are mechanically coupled, while the thumb can be driven in two positions. This allows for performing palmar grasps, where the object is grasped between all fingers and thumb, and lateral grips, in which the object is held between the lateral aspect of the index finger and the thumb. In Hannes (IIT, Rehab Technologies) and Mia hands (Prensilia, IT), the thumb, index, and middle fingers are independently controlled, while the ring and pinky are mechanically coupled to the middle finger. This enables more grasping patterns (e.g., bi-digit grip), but the maximum exerted force is lower. Finally, the maximum dexterity is reached in those hands where all fingers can be controlled individually. The challenge of shared control here is to perceive the properties of the target object and select an appropriate grasping pattern. The dexterity of the hand determines the number of options (grasps) from which the automatic controller can choose the grasping strategy \cite{mouchoux_artificial_2021}. 

Lower limb prostheses can be completely passive, essentially a spring (ankle) or damper (knee), and these simple systems are not candidates for shared control. Microprocessor-controlled prostheses, such as Genium (Ottobock), integrate sensors measuring kinematics (encoders and inertial measurement units, IMUs) and forces/contacts with the ground and use this information to adjust the joint impedance (damping level).  
Powered prostheses became commercially available recently, integrating a motor in the ankle (Empower, Ottobock) or knee (Power Knee, Össur) to provide active assistance during walking.  Here, the aim of smart control is to detect the type of terrain (e.g., stairs) and its parameters (e.g., stair height) and correspondingly set the system features (damping and stiffness) and behavior (e.g., from walking to stair climbing mode) \cite{krausz_survey_2019}.

Exoskeletons augment and restore motor functions in impaired limbs. Actively powered exoskeletons with motors are the most interesting shared control systems. Hand exoskeletons restore grasping, and upper and lower limb exoskeletons help with walking. The control challenges and strategies are similar to those used in smart prosthetic limbs.

\subsubsection{Sensory Sources}\label{sec:datasources}

To implement shared control, bionic limbs have many sensors. Most importantly, the user remains the ultimate controller and can override the system's automatic decisions with voluntary commands. This is implemented by decoding EMG signals recorded using electrodes placed on the surface of the skin of the residual limb. The signals are processed by ML methods to recognize the user's motion intention and control the device accordingly. Studies indicate that, while several channels can detect basic wrist and hand motions~\cite{mouchoux_artificial_2021}, more channels are needed to detect subtle commands like finger movements in a dexterous robotic prosthetic hand \cite{_myoplus_}. An electrode matrix with dense pads integrated into a prosthesis socket can record high-density EMG. A 64-channel matrix around the residual limb's proximal part is a realistic configuration, sampling EMG signals at 2 kHz with 16 bits of precision and leading to a data rate of 2.048 Mb/s. 

Automatic control can be implemented with computer vision. Placing cameras and depth sensors on the bionic limb helps to perceive and analyze the environment in front of the system~\cite{castro_continuous_2022}. Point-cloud analysis can detect an object in front of a prosthetic hand, determine its shape and size, and automatically preshape the prosthesis for grasping. Bionic prosthetic legs and exoskeletons can use deep learning to detect terrain and take appropriate actions \cite{krausz_survey_2019}. The sensors can also be positioned on the user \cite{mouchoux_artificial_2021}, e.g., by using smart glasses integrating miniature cameras. In this case, the system is not self-contained anymore, but the field of view is increased, improving the capabilities of the automatic controller. The camera can detect multiple objects and track the prosthesis in space, detecting user intention (target object) and hand-object interaction points. Camera configuration balances system autonomy, environmental coverage, and computational constraints.  The number of cameras, image resolution, frame rate, and codecs will determine data-rate requirements. Raw video feeds, which avoid the latency of the locally executed compression, can require rates from 73.27 Mb/s (424x240 RGB signal with 30 frames per second) to 2.98 Gb/s (1920x1080 RGB signal with 60 frames per second). 

Automatic control of bionic limbs, especially upper limb prostheses, is additionally challenged by the fact that the artificial controller does not have control over all degrees of freedom. While the controller can rotate the wrist and command the hand, the orientation of the limb in space and the approach trajectory towards the target object are selected and executed by the user, who chooses how to move the residual limb. However, this is important information for the automatic controller since the side from which to grasp the object determines the best grasping strategy. Motion sensors, like inertial measurement units (IMUs), can be used to detect motion and orientation \cite{mouchoux_artificial_2021}. They are convenient for integration into prosthesis sockets as they are miniature and low-power. Typically, an IMU integrates 3-axial accelerometers, gyroscopes, and magnetometers, and state-of-the-art units are also equipped with onboard sensor fusion estimating the three Euler angles (e.g., BNO0055, Bosch). Since normal human movements are not particularly fast, especially when using bionic limbs, the three kinematic signals can be sampled at 50 Hz with 8-bit precision, totaling the data rate of 3.6 kb/s.
          
Another modality relevant for automatic control is tactile sensing. Tactile sensors placed on the limb help assess physical attributes via direct contact with the object~\cite{kappassov_tactile_2015}, delivering critical feedback regarding interaction forces and surface characteristics at contact points. Some commercial prostheses are equipped with sensors measuring the total grasping force (robotic hands) or ground reaction force (lower limb prosthesis and exoskeletons). A higher fidelity contact information can be obtained using distributed tactile sensing. To this aim, a robotic hand or foot can be endowed with an electronic skin - a flexible sheet with an integrated network of sensing points (taxels). Such high spatial and time resolution measurements for advanced control, e.g.,  can be used to guide the closure of individual fingers of a robotic hand to achieve a stable grasp or to estimate the aspect of the foot in contact with the ground, which is critical for stability during assisted walking. Assuming a skin with 64 sensors distributed around the robotic hand/foot generating dynamic tactile signals sampled at 2 kHz for precise detection of contact timing and with the precision of 8 bits to capture proportional pressure information, the required data rate to transmit the tactile information becomes 1.024 Mb/s.

\subsubsection{Feedback interfaces}

Sensory feedback is critical for the planning and execution of movements in healthy humans.
Hence, it is important to provide artificial sensory feedback to the users of bionic limbs to close the control loop by measuring the relevant data from the sensors integrated into the limb and conveying this information to the user via a suitable interface. The most common approach is to use haptic feedback, wherein mechanical or electrical stimulation is applied to the user's skin to produce tactile sensations~\cite{dosen_prosthetic_2021}. The force between the hand and an object (hand prostheses) or between the ground and the foot (leg prostheses and exoskeletons) can be conveyed by increasing the intensity or frequency of electro- or vibrotactile stimulation, allowing the user to "feel" their bionic limbs. If the bionic limb is equipped with an electronic skin, the high-density tactile information can be transmitted to the user by delivering electrotactile stimulation through a matrix electrode. The simplest approach would be to associate each pad of the electrode with a sensor in the skin. Assuming 64 sensor skin, as mentioned in the previous section, this leads to 64 tactile feedback signals delivered to the user at 100 Hz for smooth sensations with the precision of 8 bits to convey intensity information (pressure magnitude). The total data rate for such feedback would be 51.2 kb/s. 

Shared control requires tight synchronization and collaboration between the user and the smart bionic limb. To facilitate this interaction, the feedback to the user can be enhanced by exploiting the visual channel since vision provides high throughput and fidelity. The user can be equipped with extended reality (XR) glasses (e.g., XREAL glasses, Meta Quest glasses), employed to convey not only the state of the limb (e.g., grasping force and hand position) but also the decisions of the automatic controller (e.g., detected target object, selected grasp type), allowing the user to correct the system when required \cite{mouchoux_artificial_2021}. The feedback is provided by projecting virtual objects into the real environment next to the object to be grasped (e.g., a force bar, a picture of the selected grasp), ensuring clear and simple interpretation. Assuming that commercial XR glasses are used to provide this feedback with a 4Kx4K visual field and 60 frames per second, the resulting data rate is roughly 400 Mb/s~\cite{mangiante_vr_2017}. 

\subsection{Local Unit}
It is responsible for several essential tasks:
\begin{itemize}
    \item Controller Software: The unit processes sensor data locally through tiny ML/AI algorithms, whereas more intensive processing tasks are offloaded via the access network. 
    \item Failsafe Operation: The unit runs basic functions via the local controller when network connectivity fails,  maintaining safety and lowering security risks \cite{chiariotti_future_2024}. For instance, when the connection is available, the user can control individual finger movements of a dexterous prosthesis thanks to powerful ML algorithms running on the edge, whereas, in the case of connection loss, the local controller takes over and implements simple ML methods recognizing only the stereotypical gross motions (e.g., grasp types). The system might use one of the feedback channels to alert the user about the switching from the edge to local control leading to (temporarily) reduced system capabilities, thereby ensuring explicit and graceful degradation. 
    \item Sensor Integration: The unit may connect to onboard sensors, see Section~\ref{sec:datasources}, acting as the hub for the acquisition and preconditioning of the sensor signals, relaying them for further processing to the local or edge controller.
\end{itemize}
Local control will be implemented using well-established state-of-the-art methods (e.g., pattern classification of EMG \cite{_myoplus_}) and hence, if the edge/cloud connection is not available, the advanced functions will be temporarily suspended but the connected bionic limb will still perform at the level of conventional systems.

\subsection{Wireless Access Network}
The access network connects the bionic limb to edge servers by transmitting the sensed data from the device to the edge and computed command and feedback signals from the edge and back to the device and the user, respectively.  

Apart from the indicated latency requirement on closing of the control loop in 100-125~ms, which sets the limits on the communication delays allowed in the system, in a fully fledged scenario with rich sensory (EMG, video, IMU, tactile sensing), feedback (tactile and video feedback) and control streams, the access network should support data rates in the order 100 Mb/s to 2 Gb/s in the uplink, depending on the frame resolution and rate, and the 400 Mb/s in the downlink.
Obviously, the requirements of a high-end system can be supported only by resorting to transmissions in mmWave band, while in the lower-end case, sub-6 GHz systems can be a suitable choice.
We investigate this aspect in Section~\ref{sec:evaluation}.

Access networks could also prioritize certain data streams. In the downlink, the command signals require the quickest transmission to ensure a timely reaction by the system, followed by feedback signals where larger delays can be tolerated. Similarly, in the uplink, EMG recordings may get prioritized over other sensory data as they encode explicit user intention. 
In this respect, solutions like network slicing in the access and/or the emerging frameworks for customized deployments like O-RAN~\cite{polese_understanding_2023} can facilitate the required quality of service.

Finally, it should also be noted that the use of wireless connectivity inherently poses security concerns, which should be appropriately addressed using authentication, authorization, and data encryption protocols, which should protect communication channels in the system end-to-end.

\subsection{Edge Servers}
Dedicated edge server(s) process sensor data using advanced, computationally demanding methods to calculate command and feedback signals for the bionic limb. In this online control loop,  the device-edge-device round-trip latency must be minimized. Importantly, a single server can serve multiple bionic limbs in its coverage. This can include any of the following functions already proposed in the literature as the main drivers of future smart bionics:
\begin{itemize}
    \item Advanced ML/AI: Deep learning shows promise for EMG signal decoding using classification and regression \cite{marinelli_active_2023}. A specialized ML edge server can run complex deep networks to predict a large number of movements in real-time, allowing the user to seamlessly control a dexterous prosthesis with many degrees of freedom. 
    \item Musculoskeletal Modeling:  This approach uses complex neuromusculoskeletal models to reconstruct the missing or impaired limb \cite{marinelli_active_2023}. The model leverages the EMG to estimate the motion of the virtual limb (digital twin) and generate commands for the bionic limb to recreate the virtual system's behavior.
    
    \item Context Analysis: Computer vision and contextual processing tasks are essential for implementing smart bionic limbs ~\cite{chiariotti_future_2024}. The computer vision pipeline can comprise point cloud processing  \cite{castro_continuous_2022} or deep learning to recognize important features in the environment (e.g., objects for hand prostheses and terrain types for bionic legs) and automatically select actions (e.g., grasping strategy or walking modality) \cite{krausz_survey_2019}.  The system can also continuously update and enrich the environmental model, thereby improving the control capabilities over time. 
    
    \item User feedback: The feedback to the user provides an insight into the state of the bionic limb and motion execution. This can include rendering graphical objects for augmented reality glasses or fusing and processing tactile data from the electronic skin to compute haptic stimulation profiles.

\end{itemize}

In practice, edge computing can be realized via multiple approaches, e.g., multi-access edge computing supported by cellular operators or proprietary solutions leveraging cellular non-public networks (NPN).
We also note that all computational and data storage processes in the system, both at the edge as well as in the cloud, should leverage privacy-preserving techniques, such as data anonymization or pseudonymization, to reduce privacy risks. Compliance with applicable data protection requirements, e.g., GDPR, will be an important factor in the system's design and implementation.

\subsection{Core Network}

The core network establishes the foundation for communication between edge and cloud servers. 
It can include both public Internet as well as private, customized, and optimized solutions, e.g., NPNs.

\subsection{Cloud Servers}

Cloud servers offer high-performance resources for long-term data storage and other non-time-critical operations. This infrastructure manages various functions:

\begin{itemize}
\item ML/AI Training: Cloud servers train and update AI models potentially using aggregated data from multiple prosthetic devices~\cite{chiariotti_future_2024}. This enables the training of large neural networks for complex tasks, the development of generic control models that can be quickly customized for a specific user, and the implementation of federated learning across multiple users while preserving privacy. The system can continuously improve its control algorithms based on collective user experiences through these capabilities. Once the models are trained, they can be deployed to the dedicated edge servers for online control.

\item Data Storage and Analysis: The cloud maintains extensive databases that track user activity patterns and preferences, describe typical movement patterns, store environmental maps, object models, and gather system performance measurements. This data allows for long-term monitoring of individual and aggregate usage trends, facilitating continual system development and customization.

\item Healthcare Provider Interface: Medical practitioners can remotely monitor prosthesis usage and user activity levels, allowing for early diagnosis of potential difficulties \cite{chiariotti_future_2024}. In turn, the healthcare personnel can provide guided training in users' natural settings and respond quickly to emergencies or malfunctions.

\item Over-the-Air Updates: Cloud infrastructure provides remote deployment of enhanced control algorithms, firmware, and software upgrades for prostheses, dissemination of new features and capabilities, and security vulnerability patches. 

\end{itemize}

\section{Experimental Setup and Results}
\label{sec:evaluation}

This section presents the experimental study evaluating the access network performance for edge-controlled prosthetic hand, as an example of a connected bionic limb.\footnote{The study is motivated by a recently demonstrated experimental setup for semi-autonomous hand prosthesis control~\cite{castro_continuous_2022} exploiting a wired connection between the prosthetic hand and the computer executing the control algorithm.}

\begin{figure*}[!t]
    \centering
    \includegraphics[width=0.5\linewidth]{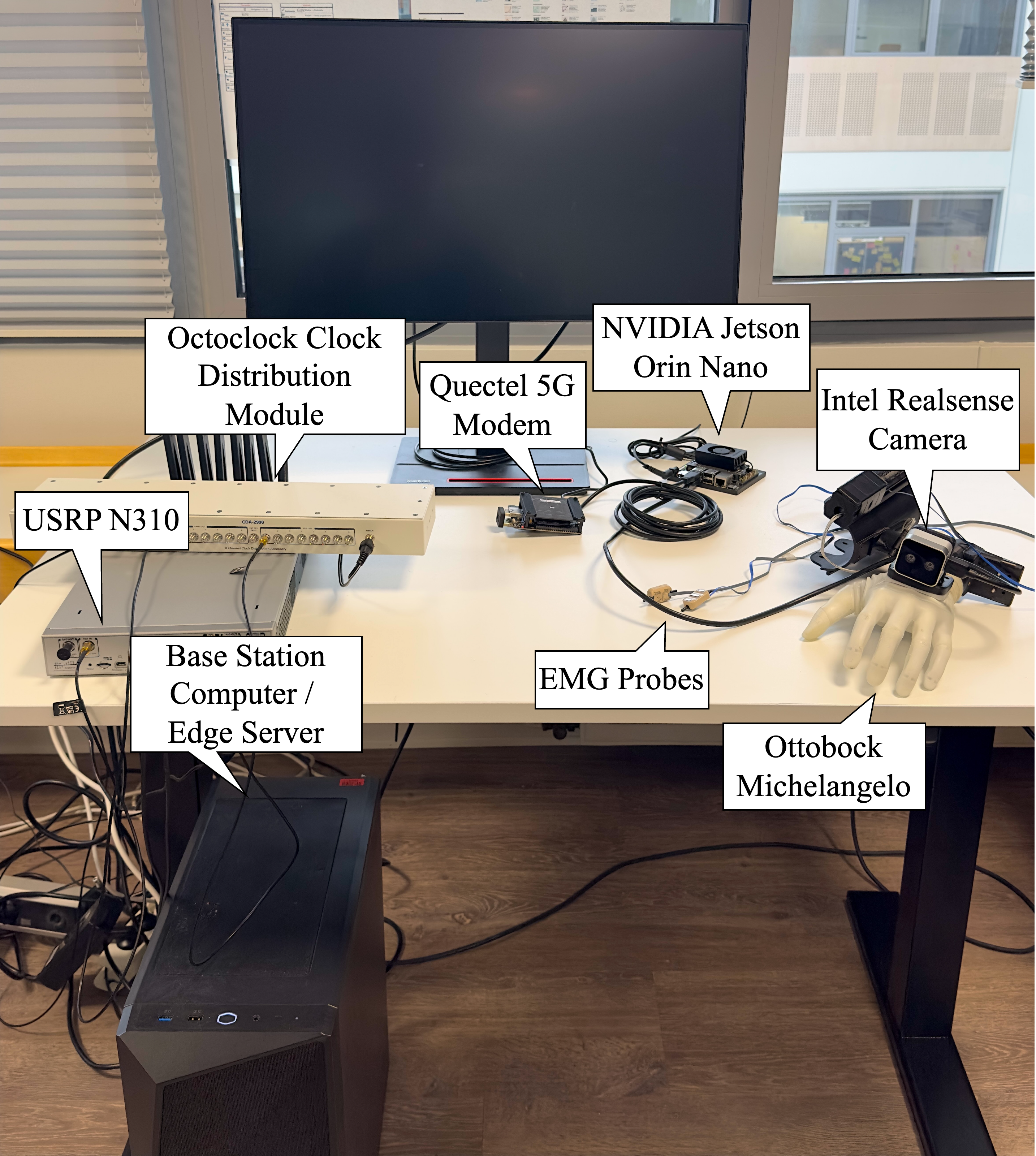}
    \caption{Experimental setup comprising a USRP N310 acting as the radio unit of the base station implemented in the desktop computer, edge server implemented in the same desktop computer, and a remotely controlled prosthetic hand with a mounted Intel Realsense camera streaming images over the local unit (Jetson device with a 5G modem).}
    \label{fig:experimental_setup}
\end{figure*}

\autoref{fig:experimental_setup} depicts a 5G testbed using USRP n310, srsRAN, and open5GS core network, operating at 4.1-4.2 GHz.
The RealSense camera mounted on the hand is connected to a Jetson device attached to Quectel 5G modem and transmits raw RGBD image signals of size 424x240 pixels that include depth information~\cite{castro_continuous_2022} (3.26 Mb in total per frame). The images are used to perform object detection at the edge server and to automatically preshape the hand based on the estimated properties of the target object. 

Table \ref{tab:network_perf} shows the data throughput and Round Trip Time (RTT) measured under 4G and 5G network configurations. 
In the 4G setup, we evaluated configurations with 10 MHz and 20 MHz bandwidths.
5G setups employed 2x2 MIMO technology, and we tested both standard and setting optimized for improved uplink performance, critical for transferring sensory data from the prosthetic device to the edge server.
The standard configuration uses default parameters with 6 downlink slots, 3 uplink slots, QAM64 modulation, and standard scheduling timings.
Our optimized settings ("opt.") featured targeted changes to the baseline parameters: we used an asymmetric TDD pattern with 7 uplink slots and only 2 downlink slots, reduced the scheduling request period from 20ms to 10ms, minimized the latency configuration of the error-correction mechanism, and enabled the QAM256 in the uplink.
The table shows that increased bandwidth and link optimization can provide tremendous improvements in the uplink data rates. 
In contrast, the differences in the average RTT performance among configurations were modest. 

The access network latency, which comprises the frame transmission time and the RTT (i.e., the processing time at the edge server is excluded) for each configuration, is shown in Table \ref{tab:network_time}; note that the downlink transmission latency is neglected in the calculation, as the payload size of command signals is within the range of a few bytes in the considered scenario.
Obviously, the uplink speed of 4G (10 MHz) is insufficient to support the latency budget of 125 ms required for seamless control of the prosthetic device.
For 4G (20 MHz), the access network latency is 95 ms, leaving about 30 ms of the latency budget for the edge processing.
On the contrary, the 5G network has the potential to provide substantially lower latencies of the uplink transmission, for instance, with the optimized 5G link with 100 MHz,  and there is ca. 85 ms of the latency budget remaining. 

In summary, the results presented in Tables~\ref{tab:network_perf} and \ref{tab:network_time} indicate that the optimized 5G scenarios can support substantial transmission rates and low latency, leaving enough bandwidth and/or latency budget for concurrent transmissions of other data streams in the system. This can be accompanied by techniques like network slicing to guarantee the QoS for particular streams.
Moreover, the remaining latency budget would also allow for data compression at the device/sensor side, thereby boosting the potential for concurrent and timely transmissions of multiple streams.

\begin{table}[!t]
\centering
\caption{Measured Network Performance}
\label{tab:network_perf}
\begin{tabular}{lrrr}
\hline
Configuration & Uplink speed & Downlink speed & Average RTT  \\
\hline
4G (10 MHz) & 22 Mb/s & 47 Mb/s & 24 ms \\
4G (20 MHz) & 48 Mb/s & 93 Mb/s & 27 ms \\
5G (60 MHz) &  60 Mb/s & 302 Mb/s & 23 ms \\
5G (100 MHz) &  107 Mb/s & 244 Mb/s & 30 ms \\
5G (60 MHz) opt. & 180 Mb/s & 99 Mb/s & 27 ms \\
5G (100 MHz) opt. & 236 Mb/s & 160 Mb/s & 27 ms \\
\hline
\end{tabular}
\end{table}
\begin{table}[!t]
\centering
\caption{Transmission time under different network configurations}
\label{tab:network_time}
\begin{tabular}{lrr}
\hline
Configuration & Frame transmission time & Access network latency \\
\hline
4G (10 MHz) & 148 ms & 172 ms\\
4G (20 MHz) & 68 ms &  95 ms \\
5G (60 MHz) &  54 ms & 77 ms  \\
5G (100 MHz)  &  30 ms & 60 ms  \\
5G (60 MHz) opt. &  18 ms & 45 ms  \\
5G (100 MHz) opt. &  14 ms & 41 ms  \\
\hline
\end{tabular}
\end{table}

\section{Discussion and Conclusions}
\label{sec:conclusion}
We presented a conceptual framework for cloud-connected bionic limbs that addresses key limitations of the current prosthetic technology, like local computational constraints and limited environmental perception. The framework features a three-tier architecture that distributes processing tasks across local, edge, and cloud controllers to maximize performance while respecting the latency required to support different functions of the smart connected bionic limbs. 

Our experimental results demonstrated the potential of 5G to provide the data rates and latencies needed to support the offloading of computing-intensive tasks.
The results emphasize that higher bandwidth configurations over the access network are essential for meeting the typical latency budget required for seamless control and intuitive user experience.  

The practical realization of cloud-connected bionic limbs is still in its infancy, and some of the key challenges to be addressed are:
\begin{itemize}
    \item  Developing lightweight AI/ML models to reduce both transmission and processing overhead.
    \item Integrating cybersecurity measures for data protection and cyber-attack prevention.
    \item Developing advanced sensor fusion techniques to improve the integration of multimodal data streams, enabling richer environmental perception and increasingly intelligent behavior.
    \item Thorough evaluation of the control performance and energy efficiency.
    \item Conducting comprehensive clinical testing to include the end-users and validate the approach in terms of user acceptance and functional improvements in daily activities.
\end{itemize}
These are left for our future work.

\section*{Acknowledgement}

The work presented in this paper was supported by the Independent Research Fund Denmark (DFF), project no. 2035-00169B ``CLIMB''.

\bibliographystyle{IEEEtran}
\bibliography{references}
\end{document}